\documentclass[preprint,12pt]{elsarticle}

\usepackage{amssymb}
\usepackage{amsthm}
\usepackage{color}

\def\Vec#1{\mbox{\boldmath$#1$}}

\journal{Annals of Physics}

\begin{document}

\begin{frontmatter}

\title{
  Rare transition event
  with self-consistent theory of large-amplitude collective motion
}

\author{Kyosuke Tsumura}
\ead{kyosuke.tsumura@fujifilm.com}
\author{Yoshitaka Maeda}
\author{Hiroyuki Watanabe}
\address{
  Analysis Technology Center,
  Research \& Development Management Headquarters,
  Fujifilm Corporation,
  Kanagawa 250-0193, Japan
}

\begin{abstract}
A numerical simulation method, based on
Dang \textit{et al.}'s
self-consistent theory of large-amplitude collective motion,
for rare transition events is presented.
The method provides a one-dimensional pathway without knowledge of the final configuration,
which includes a dynamical effect caused by not only a potential but also kinetic term.
Although it is difficult to apply the molecular dynamics simulation
to a narrow-gate potential,
the method presented
is applicable to the case.
A toy model with a high-energy barrier and/or the narrow gate
shows that while the Dang \textit{et al.} treatment
is unstable for a changing of model parameters,
our method
stable for it.
\end{abstract}


\end{frontmatter}

\section{
  Introduction
}
\label{sec:001}
\begin{figure}[tHb]
  \begin{center}
    \begin{minipage}{1.0\linewidth}
      \includegraphics[width=1.0\linewidth]{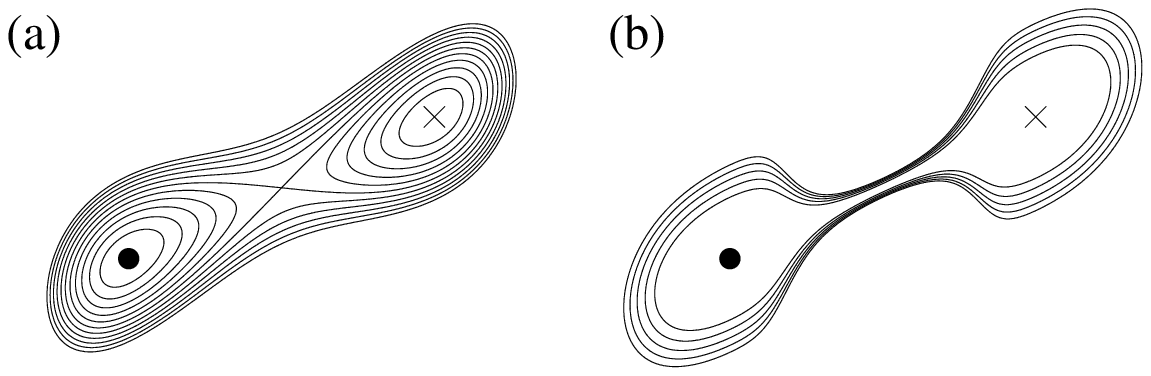}
    \end{minipage}
  \end{center}
  \caption{
    \label{fig:image}
    Schematic examples of the separations in the potential energy surface.
    (a) A high energy barrier or (b) a narrow gate
    separate the initial and final configurations
    represented by the circle and cross, respectively.
    The thin lines denote the contours of the potential energy surfaces.
  }
\end{figure}

Transition events
with long timescales
are of great importance in many fields of molecular science.
In particular,
examples in biology
range from conformational changes in proteins associated with ligand binding \cite{inducedfit}
to allosteric transitions that occur
throughout the global protein domain \cite{allosteric-0,allosteric-1}.
Their typical timescales are the order of $10^{-5}$-$1$ sec \cite{allosteric-2}.
Such rare transition events are due to
the separation between
the initial and final configurations in phase space.
The separation
is caused by the existence of
a high energy barrier or a narrow gate
in a potential energy surface;
schematic examples of such separations
are shown in Fig. \ref{fig:image}.

Numerical simulation
is an essential tool
to elucidate the rare transition event and obtain the final configuration.
There exist two kinds of approaches to the problem.
The first one is to find a pathway determined
only by the geometrical feature of the potential energy together with some criteria.
The second one is to obtain a trajectory
which respects both the potential energy and the kinetic energy.
In this work,
we employ the latter one 
because
the rare transition events are generated
by the potential and kinetic effects.

A standard method for the latter one is molecular dynamics (MD).
However,
the computational time requirement of
conventional MD
makes this method
insufficient for simulation of these rare transition events.
A number of efficient methods
that
use artificial forces
to overcome high energy barriers
have been proposed \cite{mcmd,replica,metad,taboo}.
Most of the previous methods, however,
are not efficient for rare transition events caused by a narrow gate.
This is because
the transition rate through the narrow gate
is determined by a frequency factor
and has a weak (at most power-law) dependence
on the magnitude of the artificial forces
as shown in chemical kinetics.
Both the high energy barrier and the narrow gate are
of significance for rare transition events especially in biology \cite{allosteric-2}. 

In this paper,
we describe the development of a method to simulate
rare transition events caused by not only the high energy barrier but also the narrow gate;
we are interested in transitions
between local minima in a potential energy surface.
Typical thermally fluctuating motions
are confined in space near the initial configuration;
the motions have complex multi-dimensional structures.
On the other hand,
the motion in the rare transition event
moves along a simple sharp curve for the following reasons.
For the narrow gate,
coordinates of the system are limited to specific values
as a result of the narrowness of the potential energy surface.
For the high energy barrier,
the coordinates must obtain large momenta to overcome the barrier,
and the momenta restrict the direction of motion to a specific one.
From this discussion,
the method to be developed
is one that can separate one relevant coordinate,
which parametrizes the curve,
from other coordinates orthogonal to the relevant one.
We call the coordinate a \textit{collective coordinate} and
the curve a \textit{collective path}.
Many construction methods for the collective path
have been proposed
over the past five or more decades \cite{tomonaga1,tomonaga2,marumori,nucl1-1,nucl1-2,nucl1-3,dang,walet,DKW1991,DKW,gep2}.
This approach includes
both the potential energy and the kinetic energy
as we shall see in \ref{sec:007}.

The self-consistent theory of large-amplitude collective motion
developed by Dang \textit{et al.} \cite{dang,walet,DKW1991,DKW}
is one of the powerful methods,
which is a pioneering work
in applications to molecular systems.
Central to their theory are \textit{decoupling conditions}
given by four equations whose solution gives the collective path.
Because exact decoupling is not anticipated for realistic systems,
that is, the four equations cannot have a simultaneous solution in general,
the problem becomes that of defining a constructive procedure
for the approximate collective path.
In Refs. \cite{dang,walet,DKW1991,DKW},
two approximations based on a subset of the four equations
have been proposed.

However,
which approximations in the theory of Dang \textit{et al.}
give a transition event between local minima in a potential energy surface
depends highly on the details of the potential parameters, 
as will be demonstrated in Sec. \ref{sec:002-3}.
Owing to the lack of the robustness in the previous works,
defining a unique approximation
to the decoupling conditions
and a method for evaluating the quality of it
is a central issue for development of
the collective path theory. 

A natural improvement of the approximations proposed by Dang \textit{et al.}
is a construction of the collective path
as an optimal solution to all the four decoupling condition equations
rather than a subset of them.
To manipulate this approximation and evaluate the quality of the resultant collective path,
we introduce a unique scalar quantity
that measures the deviations from the equalities in the four decoupling condition equations.
We then minimize this quantity
to formulate the proper construction method of the collective path.
The resultant method
is based on a first-order differential equation
whose solubility algorithm requires
calculation of the first and second derivatives of the potential energy at each step
for a stepwise construction of the collective path.
We show
this method
can robustly construct the collective path
even if the initial and final configurations are separated
by a high energy barrier or narrow gate
as shown in Fig. \ref{fig:image}.

This paper is organized as follows:
In Sec. \ref{sec:002},
we introduce the decoupling conditions in the theory of Dang \textit{et al.} \cite{dang,walet,DKW1991,DKW}
and demonstrate that the two approximations cannot work
as a robust construction method of the collective path.
A detailed derivation of the decoupling conditions
is presented in \ref{sec:007}.
In Sec. \ref{sec:003},
we define the scalar quantity
and then
formulate
the proper method
for the construction
of the collective path
by minimizing this quantity.
In Sec. \ref{sec:004},
we use two simple Hamiltonians to
demonstrate the ability of
our method
to construct the collective path.
We devote Sec. \ref{sec:006} to a summary and concluding remarks.

\section{
  Preliminary
}
\label{sec:002}
In this section,
we present a brief account of
the self-consistent theory of large-amplitude collective motion
developed by Dang \textit{et al.} \cite{dang,walet,DKW1991,DKW}.
First,
we present an explicit form of the decoupling conditions,
and prove that there exists no simultaneous solution to them in the generic case.
Then,
we introduce the two approximations to the decoupling conditions
that Dang \textit{et al.} \cite{dang,walet,DKW1991,DKW} have proposed.
Furthermore,
we demonstrate that
whether the two approximations can detect the collective path connecting local minima
depends highly on the shape of a potential energy surface.
We note that
the proof of the absence of the solution
and the demonstration of the lack of the robustness property
are original arguments of this work.

\subsection{
  Decoupling conditions
}
\label{sec:002-1}
Because we are interested in the applications to molecular systems,
we start with a Hamiltonian defined by
\begin{eqnarray}
  \label{eq:001}
  H = \sum_{i=1}^N \, \frac{1}{2\,m_i} \, p^2_i + V(x_1,\,\cdots,\,x_N),
\end{eqnarray}
where $N$ is the total number of degrees of freedom,
$x_i$, $p_i$, and $m_i$ denote
the coordinate, conjugate momentum, and mass of the $i$-th degree of freedom, respectively,
and $V(x_1,\,\cdots,\,x_N)$ is the potential energy.

We introduce the decoupling conditions
that define the collective path in the Hamiltonian (\ref{eq:001}).
A detailed derivation of the decoupling conditions
is presented in \ref{sec:007}.
When we write
the collective path as
\begin{eqnarray}
  \label{eq:006}
  x_i = x_i(q),
\end{eqnarray}
where $q$ denotes the collective coordinate,
the decoupling conditions to determine
the explicit $q$-dependence of $x_i$ \cite{dang,walet,DKW1991,DKW} are
\begin{eqnarray}
  \label{eq:002}
  \frac{\mathrm{d}x_i}{\mathrm{d}q} = \varphi_i/\sqrt{m_i},
\end{eqnarray}
where $\varphi_i$ are given as a solution to the following equations:
\begin{eqnarray}
  \label{eq:003}
  \sum_{j=1}^N \, (V_{,ij} - \lambda\,\delta_{ij} ) \, \varphi_j &=& 0,\\
  \label{eq:004}
  V_{,i} &=& \gamma \, \varphi_i,\\
  \label{eq:005}
  \sum_{i=1}^N \, \varphi_i \, \varphi_i &=& 1.
\end{eqnarray}
where 
\begin{eqnarray}
  V_{,i} &\equiv& \frac{1}{\sqrt{m_i}} \, \frac{\partial V(\Vec{x})}{\partial x_i},\\
  V_{,ij} &\equiv& \frac{1}{\sqrt{m_i}\,\sqrt{m_j}} \frac{\partial^2V(\Vec{x})}{\partial x_i\,\partial x_j},
\end{eqnarray}
with $\Vec{x}=(x_1,\,\cdots,\,x_N)$.
We note that
Eqs. (\ref{eq:002})-(\ref{eq:005}) show that
$\varphi_i/\sqrt{m_i}$ is a tangent vector
to the collective path $\Vec{x}=\Vec{x}(q)$ at the collective coordinate $q$,
$\varphi_i$ should be an eigenvector of the matrix $V_{,ij}$,
be proportional to the vector $V_{,i}$,
and be normalized
and that $\lambda$ and $\gamma$ denote an eigenvalue of the eigenvector $\varphi_i$
and a proportionality factor, respectively.
By solving the decoupling conditions given by Eq. (\ref{eq:002}) together with Eqs. (\ref{eq:003})-(\ref{eq:005}),
we can construct the collective path $\Vec{x} = \Vec{x}(q)$.
We use the obtained $\Vec{x}(q)$
to reduce the Hamiltonian (\ref{eq:001})
into a Hamiltonian
that governs the dynamics of the collective coordinate $q$ \cite{dang,walet,DKW1991,DKW}.

\subsection{
  Proof for absence of solution to decoupling conditions
}
\label{sec:002-2}
We emphasize that
for a generic potential energy function $V(\Vec{x})$,
there exists no $\varphi_i$ that satisfies Eqs. (\ref{eq:002})-(\ref{eq:005}),
except for the local minima or saddle points defined by $V_{,i} = 0$.
To explain this fact in an explicit manner,
first we identify $(x_i,\,\varphi_i,\,\lambda,\,\gamma)$ with would-be independent variables
and consider an infinitesimal expansion of these variables
around $(x_i,\,\varphi_i,\,\lambda,\,\gamma)$,
which satisfies Eqs. (\ref{eq:003})-(\ref{eq:005}).
Substituting
$x^\prime_i = x_i + \mathrm{d}x_i$,
$\varphi^\prime_i = \varphi_i + \mathrm{d}\varphi_i$,
$\lambda^\prime = \lambda + \mathrm{d}\lambda$,
and $\gamma^\prime = \gamma + \mathrm{d}\gamma$
into Eqs. (\ref{eq:003})-(\ref{eq:005}),
we have equations governing
$(\mathrm{d}x_i,\,\mathrm{d}\varphi_i,\,\mathrm{d}\lambda,\,\mathrm{d}\gamma)$,
which are necessary for $(x^\prime_i,\,\varphi^\prime_i,\,\lambda^\prime,\,\gamma^\prime)$
to remain the solution to Eqs. (\ref{eq:003})-(\ref{eq:005}).
Then, by dividing the equations by $\mathrm{d}q$ and incorporating Eq. (\ref{eq:002}) with them,
we reduce the decoupling conditions (\ref{eq:002})-(\ref{eq:005}) into
a compact set of equations given by
\begin{eqnarray} 
  \label{eq:new001}
  \frac{\mathrm{d}}{\mathrm{d}q}(x_i,\,\varphi_i,\,\lambda,\,\gamma)
  = (X_i/\sqrt{m_i},\,\sum_{a=1}^{N-1} \, \varphi_{\perp i}^a \, \Phi_a,\,\Lambda,\,\Gamma),
\end{eqnarray}
where
$\varphi_{\perp i}^a$ with $a=1,\,\cdots,\,N-1$ denote a complete set of the vectors orthogonal to $\varphi_i$
with the normalizations
\begin{eqnarray}
  \label{eq:new005}
  \sum_{i=1}^N \, \varphi_{\perp i}^a \, \varphi_{\perp i}^b &=& \delta^{ab},
\end{eqnarray}
and
$Y \equiv {}^t(X_i,\,\Phi_a,\,\Lambda,\,\Gamma)$ are defined as the solution to
\begin{eqnarray}
  \label{eq:new006}
  A \, Y - b = 0,
\end{eqnarray}
with
\begin{eqnarray}
  \label{eq:new007}
  A &\equiv&
{\scriptstyle 
  \left(
  \begin{array}{cccc}
    V_{,ijk}\,\varphi_k
  &
    (V_{,ij} - \lambda\,\delta_{ij})\, \varphi_{\perp j}^{a}
  &
    - \varphi_i
  &
    0
  \\
    V_{,ij}
  &
    - \gamma \, \varphi_{\perp i}^{a}
  &
    0
  &
    - \varphi_i
  \\
    \delta_{ij}
  &
    0
  &
    0
  &
    0
  \end{array}
  \right)
},\\
  \label{eq:new008}
  b &\equiv& {}^t(0,\,0,\,\varphi_i),
\end{eqnarray}
where
\begin{eqnarray}
  \label{eq:new009}
  V_{,ijk} \equiv \frac{1}{\sqrt{m_i}\,\sqrt{m_j}\,\sqrt{m_k}} \frac{\partial^3V(\Vec{x})}{\partial x_i\,\partial x_j\,\partial x_k}.
\end{eqnarray}
Here,
we have used the Einstein summation convention for the dummy indices.
We note that
the matrix $A$ is a $(3\,N)\times(2\,N+1)$ matrix.
Because
the number of the equations is obviously larger than that of the variables,
Eq. (\ref{eq:new006}) has no solution with respect to $Y = {}^t(X_i,\,\Phi_a,\,\Lambda,\,\Gamma)$ in general,
and hence the collective path $\Vec{x}(q)$ also cannot be constructed
as a solution to Eq. (\ref{eq:new001}).

We prove that
Eq. (\ref{eq:new006}) has a definite solution
of $Y = {}^t(X_i,\,\Phi_a,\,\Lambda,\,\Gamma)$
when $x_i$ agrees with $x^{\mathrm{eq}}_i$, satisfying $V_{,i}(\Vec{x}^{\mathrm{eq}}) = 0$.
We write $Y$ obtained at $x^{\mathrm{eq}}_i$ as
$Y^{\mathrm{eq}} = {}^t(X^{\mathrm{eq}}_i,\,\Phi^{\mathrm{eq}}_a,\,\Lambda^{\mathrm{eq}},\,\Gamma^{\mathrm{eq}})$.
To obtain an explicit form of $Y^{\mathrm{eq}}$,
first we construct $\varphi_i$, $\lambda$, $\gamma$, and $\varphi_{\perp i}^a$
which are necessary to build the matrix elements of $A$ and $b$ in Eq. (\ref{eq:new006}).
Substituting $V_{,i}(\Vec{x}^{\mathrm{eq}}) = 0$ into Eq. (\ref{eq:004}),
we have $\gamma = 0$.
Furthermore,
because Eq. (\ref{eq:003}) requires
$\varphi_i$ and $\lambda$ to be 
eigenvectors and eigenvalues of $V_{,ij}(\Vec{x}^{\mathrm{eq}})$, respectively,
we have $\varphi_i = \phi^{0}_i$ and $\lambda = \lambda^{0}$.
Here,
we have introduced
$\phi^{\mu}_i$ and $\lambda^{\mu}$ with $\mu=0,\,\cdots,\,N-1$
as the eigenvectors and eigenvalues of $V_{,ij}(\Vec{x}^{\mathrm{eq}})$, respectively.
We suppose that
$\lambda^{0} \ne \lambda^{a}$ for any $a=1,\,\cdots,\,N-1$.
Using the eigenvectors $\phi^{a}_i$ orthogonal to $\phi^{0}_i$,
we have $\varphi_{\perp i}^a = \phi^{a}_i$.
Then, we use
$(x_i,\,\varphi_i,\,\lambda,\,\gamma,\,\varphi_{\perp i}^a)
= (x^{\mathrm{eq}}_i,\,\phi^{0}_i,\,\lambda^{0},\,0,\,\phi_i^{a})$ to construct $A$ and $b$.
After substituting these into Eq. (\ref{eq:new006}),
we come to the solution that
\begin{eqnarray}
  \label{eq:new010}
  X^{\mathrm{eq}}_i &=& \phi^{0}_i,\\
  \label{eq:new011}
  \Phi^{\mathrm{eq}}_a &=& \frac{1}{\lambda^{0} - \lambda^{a}} \,
  V_{,ijk}(\Vec{x}^{\mathrm{eq}}) \, \phi^{0}_i \, \phi^{0}_j \, \phi_k^{a},\\
  \label{eq:new012}
  \Lambda^{\mathrm{eq}} &=& V_{,ijk}(\Vec{x}^{\mathrm{eq}}) \, \phi^{0}_i \, \phi^{0}_j \, \phi^{0}_k,\\
  \label{eq:new013}
  \Gamma^{\mathrm{eq}} &=& \lambda^{0}.
\end{eqnarray}
Indeed,
$Y = {}^t(X_i,\,\Phi_a,\,\Lambda,\,\Gamma)$ has been constructed
as a definite solution to Eq. (\ref{eq:new006}) at a stationary point.

\subsection{
  Approximations to decoupling conditions
}
\label{sec:002-3}
It is a significant task to obtain an approximate solution
to Eqs. (\ref{eq:002})-(\ref{eq:005}).
In Refs. \cite{DKW1991,DKW}, Dang \textit{et al.} proposed two methods.
One respects Eqs. (\ref{eq:003})-(\ref{eq:005}), but not (\ref{eq:002}),
which is an approximation 1 (Approx1) method.
The other takes into account 
Eqs. (\ref{eq:002}), (\ref{eq:003}), and (\ref{eq:005}), but not (\ref{eq:004}),
which we call an approximation 2 (Approx2) method.

We examine whether
the Approx1 method or the Approx2 method
can construct
the collective path connecting local minima,
by using the following two-dimensional Hamiltonian ($N=2$):
\begin{eqnarray}
  H = \frac{1}{2}\,(p_1^2 + p_2^2) + V_{\mathrm{MB}}(x_1,\,x_2).
\end{eqnarray}
Here,
$V_{\mathrm{MB}}(x_1,\,x_2)$ is the M\"uller-Brown potential \cite{muller-brown},
\begin{eqnarray}
  V_{\mathrm{MB}}(x_1,\,x_2)
  = \sum_{\alpha=1}^4 \, C^\alpha \, \exp\Bigg[
  \frac{1}{2}\,\sum_{i,j=1}^2\, (x_i - \bar{x}^\alpha_i) \, M^\alpha_{ij} \, (x_j - \bar{x}^\alpha_j) \Bigg],
\end{eqnarray}
where $(C^\alpha,\,\bar{x}^\alpha_i,\,M^\alpha_{ij})$
with $\alpha = 1,\,\cdots,\,4$ and $i,\,j=1,\,2$ denote parameters.
The parameters are set equal to \cite{muller-brown}
\begin{eqnarray}
C^\alpha &=& (-200,\,-100,\,-170,\,15),\\
\bar{x}^\alpha_1 &=& (1,\,0,\,-0.5,\,-1),\\
\bar{x}^\alpha_2 &=& (0,\,0.5,\,1.5,\,1),\\
M^\alpha_{11} &=& (-2,\,-2,\,-13,\,1.4),\\
M^\alpha_{12} &=& M^\alpha_{21} = (0,\,0,\,11,\,0.6),\\
M^\alpha_{22} &=& (-20,\,-20,\,-13,\,1.4).
\end{eqnarray}
The potential energy produced by this parameter set
is denoted as
$V_{\mathrm{MB}}^{(b)}$
and
the set modified by altering $C^3$ from $-170$ to $-120$
is denoted
$V_{\mathrm{MB}}^{(a)}$.

\begin{figure}[tHb]
  \begin{center}
    \begin{minipage}{1.0\linewidth}
      \includegraphics[width=1.00\linewidth]{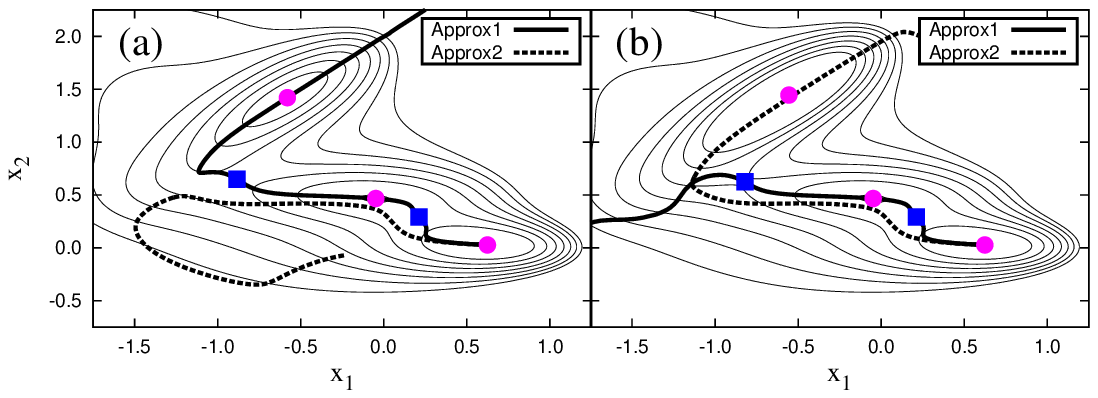}
    \end{minipage}
  \end{center}
  \caption{
    \label{fig:DKW}
    The potential energy surfaces
    made by the M\"{u}ller-Brown potential \cite{muller-brown},
    i.e.,
    (a) the modified one $V_{\mathrm{MB}}^{(a)}$
    and
    (b) the original one $V_{\mathrm{MB}}^{(b)}$,
    together with
    the collective paths
    extracted by
    the Approx1 and Approx2
    methods.
    The circles denote the local minima
    and the squares the saddle points.
    The solid (dashed)
    line represents the collective path calculated by
    the Approx1 (Approx2)
    method,
    which starts from one of the local minima.
    The thin lines denote the contours of the potential energy surfaces.
 }
\end{figure}

\begin{figure}[Hptb]
  \begin{center}
    \begin{minipage}{1.0\linewidth}
      \includegraphics[width=1.0\linewidth]{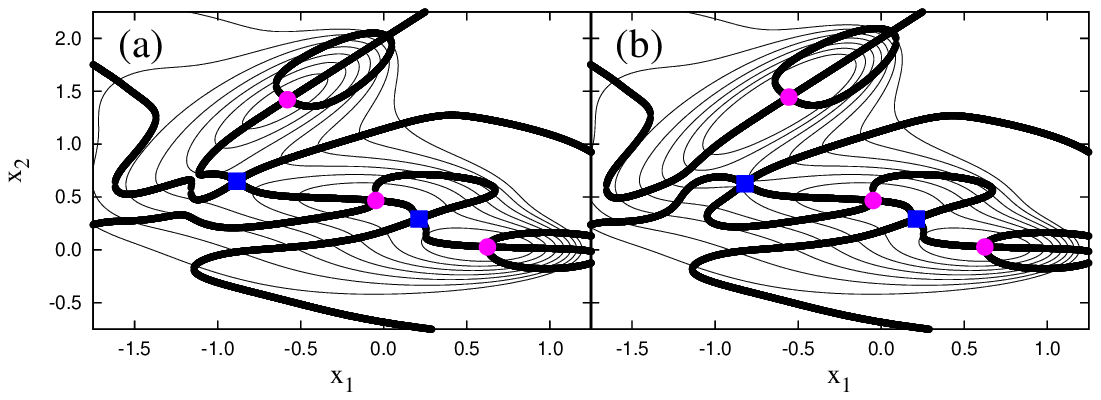}
    \end{minipage}
  \end{center}
  \caption{
    \label{fig:gep}
    All the paths derived by the Approx1 method for (a) $V_{\mathrm{MB}}^{(a)}$ and (b) $V_{\mathrm{MB}}^{(b)}$.
  }
\end{figure}

In Fig. \ref{fig:DKW},
we summarize four collective paths
that are extracted from the Hamiltonians
with $V_{\mathrm{MB}}^{(a)}$ and $V_{\mathrm{MB}}^{(b)}$
by the Approx1 and Approx2 methods.
In Refs. \cite{DKW1991,DKW},
the Approx1 method was thought to be a better method than
the Approx2 method.
For $V_{\mathrm{MB}}^{(a)}$,
indeed,
the Approx1 method can produce the collective path between local minima,
while
the Approx2 method is not the case.
For $V_{\mathrm{MB}}^{(b)}$,
however,
the approximation that extracts the transition event is not
the Approx1 method but
the Approx2 method.
Figure \ref{fig:gep} depicts
all paths based on the Approx1 method.
We note that,
in contrast to $V_{\mathrm{MB}}^{(a)}$,
$V_{\mathrm{MB}}^{(b)}$ has no single path connecting the local minima.
It turns out that
whether
the Approx1 method or the Approx2 method
can extract the transition event
sensitively depends on the shape of the potential energy surface,
and hence both
the Approx1 and Approx2
methods have not
been a robust construction method of the collective path.

\section{
  Improvement of
  approximations
  to decoupling conditions
  in self-consistent theory of large-amplitude collective motion
}
\label{sec:003}
In this section,
we develop
the numerical simulation method
that can robustly construct
the collective path connecting local minima,
which is an improvement of
the Approx1 and Approx2 methods.
A key point for the development of the method
is the construction of $Y = {}^t(X_i,\,\Phi_a,\,\Lambda,\,\Gamma)$
as an optimal solution to Eq. (\ref{eq:new006}).

By constructing the optimal solution $Y$
and substituting $Y$ into Eq. (\ref{eq:new001}),
we can obtain
a first-order differential equation
to determine $(x_i,\,\varphi_i,\,\lambda,\,\gamma)$
as a solution to an initial-value problem.

We shall construct a procedure to obtain the optimal solution to Eq. (\ref{eq:new006}). 
To this end,
first we define a vector $\Delta$, which represents
a deviation from the equality in Eq. (\ref{eq:new006}), that is,
\begin{eqnarray}
  \label{eq:new014}
  \Delta \equiv A \, Y - b.
\end{eqnarray}
Here, we treat the problem to minimize the magnitude of $\Delta$ 
to derive an equation for the construction of the optimal $Y$.

\subsection{
  Measure of deviation from exact decoupling
}
We introduce three new vectors $d_i$, $e_i$, and $f_i$
to represent $\Delta$ as a form of the modification into
$\varphi_i$ in the elements of $A$ and $b$.
Indeed,
by parameterizing $\Delta$ as
\begin{eqnarray}
\Delta =
  {\scriptstyle 
  \left(
  \begin{array}{c}
    G_{ij} \, (d_j-\varphi_j)
  \\
    \Gamma \, (e_i-\varphi_i)
  \\
    f_i-\varphi_i
  \end{array}
  \right)  
  },
\end{eqnarray}
with
\begin{eqnarray}
  G_{ij} \equiv \Lambda \, \delta_{ij} - V_{,ijk}\,X_k,
\end{eqnarray}
we can convert Eq. (\ref{eq:new014}) into the same form as Eq. (\ref{eq:new006}):
\begin{eqnarray}
  \label{eq:new015}
  \tilde{A} \, Y - \tilde{b} = 0,
\end{eqnarray}
where
\begin{eqnarray}
  \label{eq:new016}
  \tilde{A} &\equiv&
{\scriptstyle 
  \left(
  \begin{array}{cccc}
    V_{,ijk}\,d_k
  &
    (V_{,ij} - \lambda\,\delta_{ij})\, \varphi_{\perp j}^{a}
  &
    - d_i
  &
    0
  \\
    V_{,ij}
  &
    - \gamma \, \varphi_{\perp i}^{a}
  &
    0
  &
    - e_i
  \\
    \delta_{ij}
  &
    0
  &
    0
  &
    0
  \end{array}
  \right)
},\\
  \label{eq:new017}
  \tilde{b} &\equiv& {}^t(0,\,0,\,f_i).
\end{eqnarray}
Here, $\tilde{A}$ and $\tilde{b}$ are obtained by the replacement of $\varphi_i$
in the first and second rows of $A$
and the third row of $b$
to $d_i$, $e_i$, and $f_i$, respectively. 
Because the condition $d_i = e_i = f_i = \varphi_i$ reproduces the exact decoupling conditions,
we can reach the following natural definition of the scalar quantity
that measures the deviation from exact decoupling:
\begin{eqnarray}
  \label{eq:new018}
  Q = \sum_{i=1}^N\, \Big(w_d \, (d_i-\varphi_i)^2
  + w_e \, (e_i-\varphi_i)^2 + w_f \, (f_i-\varphi_i)^2\Big),
\end{eqnarray}
where $w_d$, $w_e$, and $w_f$ denote weight parameters
to determine which decoupling conditions (\ref{eq:002})-(\ref{eq:004}) are respected.
Because we are interested in the case where
all of the decoupling conditions (\ref{eq:002})-(\ref{eq:004}) are respected equivalently,
we must set
\begin{eqnarray}
  \label{eq:new019}
  (w_d,\,w_e,\,w_f)=(1,\,1,\,1).
\end{eqnarray}
By definition,
the set of $(w_d,\,w_e,\,w_f)=(1,\,1,\,0)$ corresponds to the Approx1 method,
while the set of $(w_d,\,w_e,\,w_f)=(1,\,0,\,1)$ leads to the Approx2 method.

To make clear a geometrical meaning of $Q$,
we rewrite $Q$ as
\begin{eqnarray}
  \label{eq:new020}
  Q = {}^t\Delta \, C \, \Delta = {}^t(A\,Y-b) \, C \, (A\,Y-b).
\end{eqnarray}
This representation means that
$Q$ is identical to the norm, i.e., the square of the length of the vector $\Delta = A\,Y-b$
measured on the metric $C$ given by
\begin{eqnarray}
  \label{eq:new021}
  C \equiv
  {\scriptstyle 
  \left(
  \begin{array}{ccc}
    [{}^t(G^{-1})\,G^{-1}]_{ij}
  &
    0
  &
    0
  \\
    0
  &
    \Gamma^{-2} \, \delta_{ij}
  &
    0
  \\
    0
  &
    0
  &
    \delta_{ij}
  \end{array}
  \right)
  }.
\end{eqnarray}
It is obvious that
$C$ has a dependence on $(X_i,\,\Lambda,\,\Gamma)$ that is not yet determined.

As shown in Eqs. (\ref{eq:new010})-(\ref{eq:new013}),
$Y = {}^t(X_i,\,\Phi_a,\,\Lambda,\,\Gamma)$ can be obtained
without reference to $C$
at the local minima and saddle points $x_i^\mathrm{eq}$.
If we take $x_i^\mathrm{eq}$ as a starting point to detect the collective path,
we can carry out a straightforward construction of $C$
by substituting $(X_i,\,\Lambda,\,\Gamma)$
in Eqs. (\ref{eq:new010}), (\ref{eq:new012}), and (\ref{eq:new013}) into Eq. (\ref{eq:new021}).
Here, we refer to 
$C$ constructed at the starting point $x_i^\mathrm{eq}$
as $C^{\mathrm{eq}}$.

In this work,
we use this metric $C^{\mathrm{eq}}$ as a substitute for $C$
in the course of the detection of the collective path;
\begin{eqnarray}
  \label{eq:new023}
  C = C^{\mathrm{eq}}.
\end{eqnarray}
Thanks to the metric $C$ being constant,
$Q$ is a function bilinear with respect to $Y$, which is easily minimized.
It is left to future work
to carry out the nonlinear optimization of $Q$ with respect to $Y$
by taking into account the $Y$-dependence in $C$.

\subsection{
  Basic equation of
  proper construction method of collective path
}
By applying the minimization (stationary) condition
$\partial Q/\partial Y = 0$
to $Q$ with $C=C^{\mathrm{eq}}$,
we arrive at
\begin{eqnarray}
  \label{eq:new024}
  ({}^tA\,C^{\mathrm{eq}}\,A)\,Y - {}^tA\,C^{\mathrm{eq}}\,b = 0.
\end{eqnarray}
We note that
in contrast to Eq. (\ref{eq:new006}),
Eq. (\ref{eq:new024}) can have a definite solution in general
because ${}^tA\,C^{\mathrm{eq}}\,A$ is a $(2\,N+1)\times(2\,N+1)$ matrix
and the number of the equations agrees with that of the variables.
We stress that
Eq. (\ref{eq:new001})
combined with $Y$ that is a solution to Eq. (\ref{eq:new024})
is the main equation
for the construction of the collective path,
which is one of the results in this work.

A few remarks are in order here:
(i) The basic equation of our method
given by the set of Eqs. (\ref{eq:new001}) and (\ref{eq:new024})
is a first-order differential equation with respect to $(x_i,\,\varphi_i,\,\lambda,\,\gamma)$,
which can be solved with a single iteration
for the stepwise construction of the collective path
by repeating the calculation of the elements of the matrix $A$.
In fact,
we can construct $A$ without an iteration procedure,
because the elements of $A$ are given by
only the first, second, and third derivatives of the potential energy 
with respect to the coordinate variables,
which can be calculated at a given structure of the system.
Furthermore,
we note that 
the elements associated with the third derivatives of $V(\Vec{x})$ in the matrix $A$ in Eq. (\ref{eq:new024})
can be calculated in the same computational cost as that in the case of the second derivative terms
because
the third derivative terms enter into $A$ as a form of $V_{,ijk}\,\varphi_k$.
Indeed,
we can construct $V_{,ijk}\,\varphi_k$
from the two second derivatives
on the basis of the formula
\begin{eqnarray}
  \label{eq:020}
  V_{,ijk}(\Vec{x})\,\varphi_k
  \sim \frac{V_{,ij}(\Vec{x}+\Vec{n} \, \Delta x)
          -
          V_{,ij}(\Vec{x}-\Vec{n} \, \Delta x)
  }{2\,\Delta x},
\end{eqnarray}
with $\Vec{n} \equiv (\varphi_1/\sqrt{m_1},\,\cdots,\,\varphi_N/\sqrt{m_N})$.
Here, $\Delta x$ denotes an infinitesimal value.
We note that
the calculations of the two second derivatives can be run in parallel.
Thanks to Eq. (\ref{eq:020}),
the resultant method
requires only the numerical calculation of the first and second derivatives
in each step of the single iteration.
This property reduces the computational costs considerably.
(ii) As mentioned in the final paragraph in Sec. \ref{sec:002-2},
we suppose that
the initial value of $(x_i,\,\varphi_i,\,\lambda,\,\gamma)$ is set equal to
$(x^{\mathrm{eq}}_i,\,\phi^{0}_i,\,\lambda^{0},\,0)$.
Here,
we present an account of the eigenvalue $\lambda^{0}$ and eigenvector $\phi^{0}_i$.
In this work,
we adopt the smallest eigenvalue as $\lambda^{0}$.
This is because
we are interested in the rare transition event of molecular systems
where
the amplitude of the collective coordinate
is large compared to those of the other coordinates,
which can be translated to the requirement to use the smallest eigenvalues of $V_{,ij}$.
We note that
the direction of the construction of the collective path from the initial configuration
depends on the sign of $\phi^{0}_i$,
because $\phi^{0}_i$ gives the step from the initial structure to the next one.

\section{
  Applications of
  self-consistent theory of large-amplitude collective motion
  to simple Hamiltonians
}
\label{sec:004}
In this section,
first we examine
our method's ability
to construct the collective path connecting local minima.
Then,
we elucidate the ability of
the method to extract the rare transition event
by using potential energy surfaces
with either a high energy barrier or a narrow gate
to separate the two local minima as shown in Fig. \ref{fig:image}.

\subsection{
  M\"uller-Brown potential
}
\label{sec:004-1}
To examine the robustness property of
the numerical simulation method based on the set of Eqs. (\ref{eq:new001}) and (\ref{eq:new024}),
we utilize the Hamiltonians with the M\"uller-Brown potential \cite{muller-brown},
which are the same as those presented in Fig. \ref{fig:DKW}.

\begin{figure}[tHb]
  \begin{center}
    \begin{minipage}{1.0\linewidth}
      \includegraphics[width=1.0\linewidth]{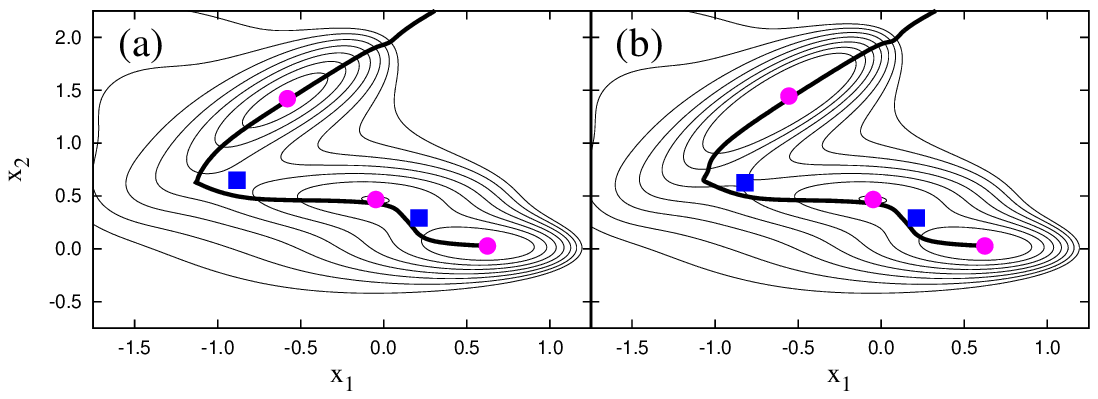}
    \end{minipage}
  \end{center}
  \caption{
    \label{fig:thiswork}
    The collective paths
    constructed by the method
    based on the set of Eqs. (\ref{eq:new001}) and (\ref{eq:new024}).
    The potential energy surfaces
    are the same as those in Fig. \ref{fig:DKW},
    and the solid line represents the collective path.
  }
\end{figure}

In Fig. \ref{fig:thiswork},
we show the resultant collective paths.
Both of them are shown as trajectories
that connect almost all the local minima and saddle points.
As that it is not the case
in the collective paths obtained with
the Approx1 and Approx2 methods,
shown in Fig. \ref{fig:DKW},
we conclude that
our method is a robust construction method
of the collective path describing the transition between local minima,
in contrast to
the Approx1 and Approx2 methods.

We now make one remark here.
The collective paths depicted in Fig. \ref{fig:thiswork}
pass close to but not exactly through the saddle points.
This is a critical problem
for the studies of minimum energy path \cite{fukui,neb}.
However,
the path studied in this paper is the one
including not only the potential energy but also the kinetic energy.
In this case,
that the path passes exactly through the saddle points is
not necessarily required.
A further examination of this point is a future problem.

\subsection{
  Barrier and gate potential
}
\label{sec:004-2}
We consider a Hamiltonian given by
\begin{eqnarray}
  H = \frac{1}{2}\,(p_1^2 + p_2^2) + V_{\mathrm{BG}}(x_1,\,x_2).
\end{eqnarray}
Here, $V_{\mathrm{BG}}(x_1,\,x_2)$ denotes the barrier and gate potential defined by
\begin{eqnarray}
  \label{eq:BG}
  V_{\mathrm{BG}}(x_1,\,x_2)
  = -\frac{1}{2}\,a\,x^2_1+\frac{1}{4}\,b\,x^4_1
  +(x_2-c\,x_1)^2 \,
  \Big[ \alpha\,x^2_1 + \beta\,\mathrm{e}^{-\gamma\,x^2_1} + \delta \Big],
\end{eqnarray}
with $(a,\,b,\,c,\,\alpha,\,\beta,\,\gamma,\,\delta)$ being parameters.
The parameters $(a,\,b,\,c)$ determine locations and energy values of stationary points.
Indeed,
this potential energy surface has two local minima and one saddle point
located at $(\pm \sqrt{a/b},\,\pm c\,\sqrt{a/b})$ and $(0,\,0)$, respectively,
and the activation energy in the transition between the two local minima is given by $a^2/(4\,b)$.
The parameters $(\alpha,\,\beta,\,\gamma,\,\delta)$ determine
the narrowness of the pathway connecting the local minima and the saddle point.

\begin{figure}[Hptb]
  \begin{center}
    \begin{minipage}{1.0\linewidth}
      \includegraphics[width=1.0\linewidth]{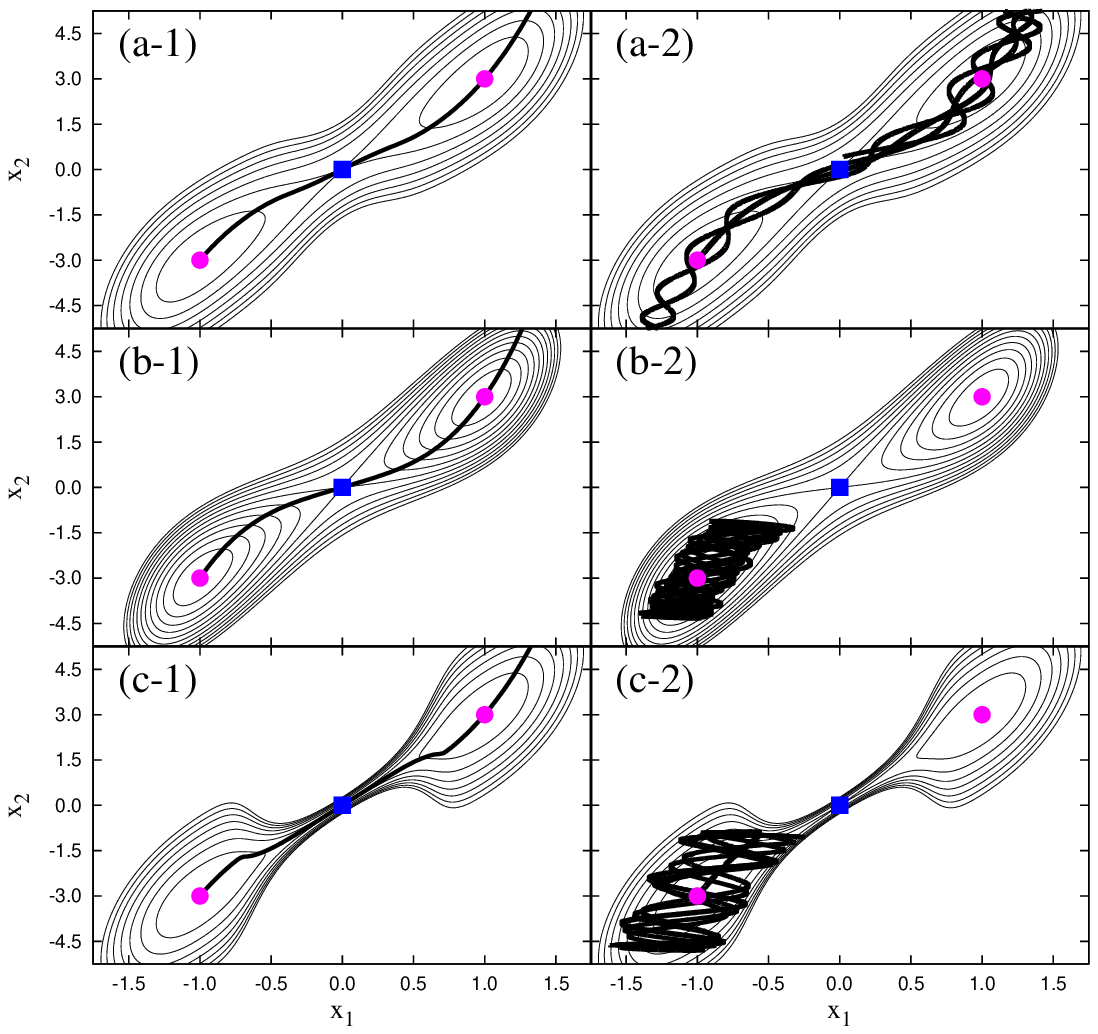}
    \end{minipage}
  \end{center}
  \caption{
    \label{fig:bgpotential}
    The collective paths obtained by
    the method developed in this work
    and the trajectories obtained by the MD method
    on the $V_{\mathrm{BG}}^{(a)}$, $V_{\mathrm{BG}}^{(b)}$, and $V_{\mathrm{BG}}^{(c)}$ potential energy surfaces.
    The solid lines in (a-1) and (a-2) ((b-1) and (b-2) or (c-1) and (c-2)) show
    the collective path
    and the MD trajectory derived from 
    $V_{\mathrm{BG}}^{(a)}$ ($V_{\mathrm{BG}}^{(b)}$ or $V_{\mathrm{BG}}^{(c)}$), respectively.
    In (a-1) and (a-2) ((b-1) and (b-2) or (c-1) and (c-2)),
    the thin lines denote the contours of $V_{\mathrm{BG}}^{(a)}$ ($V_{\mathrm{BG}}^{(b)}$ or $V_{\mathrm{BG}}^{(c)}$),
    the circles the local minima,
    and the squares the saddle point.
  }
\end{figure}

To show the ability of
the method developed to simulate the rare transition event,
we introduce three potential energy surfaces:
a potential energy surface containing neither a high energy barrier nor a narrow gate $V_{\mathrm{BG}}^{(a)}$,
a surface with a high energy barrier $V_{\mathrm{BG}}^{(b)}$,
and a surface with a narrow gate $V_{\mathrm{BG}}^{(c)}$.
The parameters
$(a,\,b,\,c,\,\alpha,\,\beta,\,\gamma,\,\delta)$
which create $V_{\mathrm{BG}}^{(a)}$, $V_{\mathrm{BG}}^{(b)}$, and $V_{\mathrm{BG}}^{(c)}$
read
$(80,\,80,\,3,\,0.0045,\,10,\,10,\,1)$,
$(240,\,240,\,3,\,0,\,0,\,10,\,20)$,
and $(80,\,80,\,3,\,0.31,\,690,\,10,\,9.7)$, respectively.
All the potential energies have the same locations of the local minima as $(\pm 1,\,\pm 3)$.

In Fig. \ref{fig:bgpotential},
we show the collective paths
corresponding to $V_{\mathrm{BG}}^{(a)}$, $V_{\mathrm{BG}}^{(b)}$, and $V_{\mathrm{BG}}^{(c)}$
together with trajectories obtained by the conventional MD method.
We find that 
the collective paths
provide pathways connecting the two local minima
even if a high energy barrier or narrow gate exists between the two local minima.
Conversely,
the MD method applied to $V_{\mathrm{BG}}^{(a)}$
can provide a trajectory near the collective path,
but the trajectories obtained for the $V_{\mathrm{BG}}^{(b)}$ and $V_{\mathrm{BG}}^{(c)}$ cases
are seen to be trapped near one local minimum and cannot reach the other local minimum.
We note that
the qualitatively same results are obtained
with less symmetric potentials, e.g., $V_{\mathrm{BG}}(x_1,\,x_2) + x_1^3$;
in \ref{sec:008},
the results for the potentials are discussed.
We stress that
in contrast to the MD method,
our method
works well as a simulation technique for the rare transition event.

\section{
  Summary and concluding remarks
}
\label{sec:006}
Basic notion adopted in this work
is that
simulation of rare transition events
can reduce to
construction of a collective path
given as a simple sharp curve
along which only one coordinate increases
if the initial coordinates are adequately chosen.
To detect the collective path,
we have developed a method
to separate a collective coordinate,
which parametrizes the collective path,
from the other coordinates orthogonal to the collective coordinate. 
The method has been formulated by constructing the collective path
as an optimal solution to the decoupling conditions
in the self-consistent theory of large-amplitude collective motion
proposed by Dang \textit{et al.} \cite{dang,walet,DKW1991,DKW}.
A basic equation of the method
is given by the set of Eqs. (\ref{eq:new001}) and (\ref{eq:new024}).
We have found that
the equation is a first-order differential equation
that determines the collective path as a solution to an initial-value problem
from an arbitrary point in the collective path
with the use of
the first and second derivatives of the potential energy with respect to the atomic coordinates.
By using the M\"{u}ller-Brown potential \cite{muller-brown},
we have demonstrated that
the method,
which is an improvement of the theory of Dang \textit{et al.},
can robustly construct the collective path
connecting the local minima in the potential energy surface.  
We have used a simple potential energy surface
to show that
our method uses only the initial configuration
to construct a pathway to the final configuration,
even if the configurations are separated by a high energy barrier and/or a narrow gate.

We consider that
our study can stimulate discussions about the rare transition events in various systems of interest.
In fact,
the method presented in this work
can be applied to simulate the rare transition events
of generic molecular systems whose potential energy surfaces are well-defined. 
For example,
by using a potential energy for biomolecules, e.g., the generalized amber force field \cite{gaff},
we will demonstrate \cite{tsumura2} that
our method
can be applied to describe
allosteric transitions \cite{allosteric-0,allosteric-1}
which are typical rare transition events for protein systems.
Furthermore,
if we utilize potential energies calculated numerically using computer programs
developed on the basis of quantum chemistry,
we can describe a chemical reaction as a rare transition event.
In fact,
we will show \cite{tsumura0} that
combining this method with the density functional theory \cite{dft1}
allows simulation of
an intermolecular proton transfer reaction,
i.e., molecular dynamics with quantum tunneling.
We will also show \cite{tsumura1} that
our method in combination with time-dependent density functional theory \cite{TDDFT}
can be applied to simulate photochemical reactions, i.e.,
the dynamics of molecular systems in electronic excited states.

Finally,
we remark again
the following study
that should be reported in the near future.
We apply the non-linear optimization to Eq. (\ref{eq:new020})
to determine the collective path
in a full consistent manner
and compare it with the collective path obtained in the approximation (\ref{eq:new023}).

\section*{Acknowledgments}
The authors are grateful to Dr. Furuya for his encouragement and support.

\appendix

\section{
  Detailed derivation of decoupling conditions
}
\label{sec:007}
In this section,
we present a review of the derivation of the decoupling conditions
in the self-consistent theory of large-amplitude collective motion
developed by Dang \textit{et al.} \cite{dang,walet,DKW1991,DKW}.
We focus on the collective path parametrized by one collective coordinate,
although Dang \textit{et al.} \cite{dang,walet,DKW1991,DKW} have presented
a construction method of a multi-dimensional collective path,
i.e., the so-called collective surface.
Basically, we obey the notation used in Refs. \cite{DKW1991,DKW} in this section.
In the final stage,
we convert the derived decoupling conditions
into those introduced in Sec. \ref{sec:002-1}.

We treat an $N$-body system whose Hamiltonian is
\begin{eqnarray}
  \label{eq:app001}
  H = \frac{1}{2}\,\pi_\alpha\,B^{\alpha\beta}\,\pi_\beta + V(\xi^1,\,\cdots,\,\xi^N).
\end{eqnarray}
Here,
$\alpha$ and $\beta$ denote indices that vary from $1$ to $N$,
$\xi^\alpha$ and $\pi_\alpha$ are a coordinate and its conjugate momentum of the $\alpha$-th degree of freedom, respectively,
$B^{\alpha\beta}$ is a reciprocal mass tensor,
and $V(\xi^1,\,\cdots,\,\xi^N)$ is a potential energy.
We suppose that $B^{\alpha\beta}$ is diagonalized as
\begin{eqnarray}
  \label{eq:app002}
  B^{\alpha\beta} = \mathrm{diag}(1/m_1,\,\cdots,\,1/m_N),
\end{eqnarray}
with $m_\alpha$ being the mass of the $\alpha$-th degree of freedom.

We use a canonical transformation
to obtain conditions
that define the collective path.
In the canonical transformation,
we consider mappings of the form
\begin{eqnarray}
  \label{eq:app003}
  \xi^\alpha = g^\alpha(q^1,\,\cdots,\,q^N) \equiv g^\alpha(q),
\end{eqnarray}
whose inverse relations are given by
\begin{eqnarray}
  \label{eq:app004}
  q^\mu = f^\mu(\xi^1,\,\cdots,\,\xi^N) \equiv f^\mu(\xi).
\end{eqnarray}
Using a chain rule relation,
we find that $g^\alpha(q)$ and $f^\mu(\xi)$ satisfy the following relations:
\begin{eqnarray}
  \label{eq:app005}
  \delta^\alpha_{\,\,\,\beta} &=& \partial \xi^\alpha/\partial \xi^\beta
  = g^\alpha_{,\,\mu}(q)\,f^\mu_{,\,\beta}(\xi),\\
  \label{eq:app006}
  \delta^\mu_{\,\,\,\nu} &=&
  \partial q^\mu/\partial q^\nu = f^\mu_{,\,\alpha}(\xi)\,g^\alpha_{,\,\nu}(q),
\end{eqnarray}
where the comma indicates a partial derivative;
$F_{,\,\alpha}(\xi) \equiv \partial F(\xi)/\partial \xi^\alpha$
and
$G_{,\,\mu}(q) \equiv \partial G(q)/\partial q^\mu$.
The indices $\alpha,\,\beta,\,\cdots$ represent those of the initial coordinates,
and $\mu,\,\nu,\,\cdots$ those of the final coordinates,
although each set has the same range $1,\,\cdots,\,N$.
The momentum $p_\mu$ conjugate to $q^\mu$ is given by
\begin{eqnarray}
  \label{eq:app007}
  p_\mu = g^\alpha_{,\,\mu}(q)\,\pi_\alpha,
\end{eqnarray}
with inverse
\begin{eqnarray}
  \label{eq:app008}
  \pi_\alpha = f^\mu_{,\,\alpha}(\xi)\,p_\mu.
\end{eqnarray}

The Hamiltonian as described with the canonical coordinates $(q^\mu,\,p_\mu)$ is
\begin{eqnarray}
  \label{eq:app009}
  H = \frac{1}{2}\,p_\mu\,\bar{B}^{\mu\nu}(q)\,p_\nu + \bar{V}(q).
\end{eqnarray}
Here,
$\bar{B}^{\mu\nu}(q)$ is an inverse matrix of $\bar{B}^{-1}_{\mu\nu}(q)$ given by
\begin{eqnarray}
  \label{eq:app010}
  \bar{B}^{-1}_{\mu\nu}(q) \equiv
  g^\alpha_{,\,\mu}(q)\,B^{-1}_{\alpha\beta}\,g^\beta_{,\,\nu}(q),
\end{eqnarray}
and $\bar{V}(q)$ is the potential energy written by $q^\mu$ as
\begin{eqnarray}
  \label{eq:app011}
  \bar{V}(q) \equiv V(\xi^1=g^1(q),\,\cdots,\,\xi^N=g^N(q)).
\end{eqnarray}
Equations of motion derived from the Hamiltonian (\ref{eq:app009}) are
\begin{eqnarray}
  \label{eq:app012}
  \dot{q}^\mu &=& \bar{B}^{\mu\nu}(q) \, p_\nu,\\
  \label{eq:app013}
  \dot{p}_\mu &=&
  - \frac{1}{2}\,p_\nu\,\bar{B}^{\nu\rho}_{,\,\mu}(q)\,p_\rho - \bar{V}_{,\,\mu}(q).
\end{eqnarray}

In the new canonical coordinates,
we identify the collective path
as follows:
We divide the set of $(q^\mu,\,p_\mu)$ into two subsets, $(q^1,\,p_1)$ and $(q^a,\,p_a)$ with $a = 2,\,\cdots,\,N$,
and suppose this division to ensure
that if at time $t = 0$ both $q^a = 0$ and $p_a = 0$,
then $q^a(t) = 0$ and $p_a(t) = 0$.
The time evolution of $q^1$ and $p_1$ in the new canonical coordinates can be converted into
that of $\xi^\alpha$ and $\pi_\alpha$ which takes place on a two-dimensional subspace
\begin{eqnarray}
  \label{eq:app014}
  \xi^\alpha &=& \hat{\xi}^\alpha \equiv g^\alpha(\hat{q}),\\
  \label{eq:app015}
  \pi_\alpha &=& \hat{\pi}_\alpha \equiv f^\mu_{,\,\alpha}(\hat{\xi})\,\hat{p}_\mu,
\end{eqnarray}
with
\begin{eqnarray}
  \label{eq:app016}
  \hat{q}^\mu &\equiv& (q^1,\,0,\,\cdots,\,0),\\
  \label{eq:app017}
  \hat{p}_\mu &\equiv& (p_1,\,0,\,\cdots,\,0).
\end{eqnarray}
We note that
$q^1$ is the collective coordinate
and $\hat{\xi}^\alpha$ is identical to the collective path parametrized by $q^1$.

The requirements $(\dot{q}^a,\,\dot{p}_a)=(0,\,0)$ can be compatible with the requirements $(q^a,\,p_a)=(0,\,0)$
only if the equations
\begin{eqnarray}
  \label{eq:app018}
  0 &=& \bar{B}^{a1}(\hat{q}) \, p_1,\\
  \label{eq:app019}
  0 &=&
  - \frac{1}{2}\,p_1\,\bar{B}^{11}_{,\,a}(\hat{q})\,p_1 - \bar{V}_{,\,a}(\hat{q}),
\end{eqnarray}
are satisfied,
as one sees from Eqs. (\ref{eq:app012}) and (\ref{eq:app013}).
Equations (\ref{eq:app018}) and (\ref{eq:app019})
are equivalent to three conditions
\begin{eqnarray}
  \label{eq:app020}
  \bar{B}^{a1}(\hat{q}) &=& 0,\\
  \label{eq:app021}
  \bar{V}_{,\,a}(\hat{q}) &=& 0,\\
  \label{eq:app022}
  \bar{B}^{11}_{,\,a}(\hat{q}) &=& 0.
\end{eqnarray}
Here,
we have assumed that $p^1$ is not a constant of the motion.
We emphasize that
these equations determine an explicit form of
$\hat{\xi}^\alpha$.
In fact,
if $g^\alpha_{,\,1}(\hat{q})$ is obtained as a solution to Eqs. (\ref{eq:app020})-(\ref{eq:app022}),
we can construct $\hat{\xi}^\alpha=\hat{\xi}^\alpha(q^1)$
by solving a differential equation
\begin{eqnarray}
  \label{eq:app023}
  \frac{\mathrm{d}}{\mathrm{d}q^1}\hat{\xi}^\alpha
  = g^\alpha_{,\,1}(\hat{q}),
\end{eqnarray}
which has been derived
from the definition (\ref{eq:app014}).

From here on,
we convert Eqs. (\ref{eq:app020})-(\ref{eq:app022})
into convenient forms to obtain an explicit form of $g^\alpha_{,\,1}(\hat{q})$. 
For this purpose,
first we introduce two scalar quantities
\begin{eqnarray}
  \label{eq:app024}
  X^{(0)}(\xi) &\equiv& V(\xi)\nonumber\\
  &=& \bar{V}(q) \equiv \bar{X}^{(0)}(q),\\
  \label{eq:app025}
  X^{(1)}(\xi) &\equiv&
  V_{,\,\alpha}(\xi)\,B^{\alpha\beta}\,V_{,\,\beta}(\xi)\nonumber\\
  &=&\bar{V}_{,\,\mu}(q)\,\bar{B}^{\mu\nu}(q)\,\bar{V}_{,\,\nu}(q)
  \equiv
  \bar{X}^{(1)}(q).
\end{eqnarray}
These quantities
have an important property given by
\begin{eqnarray}
  \label{eq:app026}
  \bar{X}^{(\sigma)}_{,\,a}(\hat{q}) = 0,\,\,\,\sigma=0,\,1.
\end{eqnarray}
Here, we present a proof for the above equations.
The proof for the case of $\sigma=0$ 
is trivial as follows:
\begin{eqnarray}
  \label{eq:app027}
  \bar{X}^{(0)}_{,\,a}(\hat{q}) = \bar{V}_{,\,a}(\hat{q}) = 0,
\end{eqnarray}
which is just Eq. (\ref{eq:app021}).
In the case of $\sigma=1$,
we simply compute
\begin{eqnarray}
  \label{eq:app028}
  \bar{X}^{(1)}_{,\,a}(\hat{q}) &=& 2\,\bar{X}^{(0)}_{,\,\mu a}(\hat{q})\,
  \bar{X}^{(0)}_{,\,\nu}(\hat{q})\,\bar{B}^{\mu\nu}(\hat{q})
  +\bar{X}^{(0)}_{,\,\mu}(\hat{q})\,\bar{X}^{(0)}_{,\,\nu}(\hat{q})\,
  \bar{B}^{\mu\nu}_{,\,a}(\hat{q})\nonumber\\
  &=& 2\,\bar{X}^{(0)}_{,\,b a}(\hat{q})\,
  \bar{X}^{(0)}_{,\,1}(\hat{q})\,\bar{B}^{b1}(\hat{q})
  + 2\,\bar{X}^{(0)}_{,\,1 a}(\hat{q})\,
  \bar{X}^{(0)}_{,\,1}(\hat{q})\,\bar{B}^{11}(\hat{q})\nonumber\\
  &&{}+\bar{X}^{(0)}_{,\,1}(\hat{q})\,\bar{X}^{(0)}_{,\,1}(\hat{q})
  \,\bar{B}^{11}_{,\,a}(\hat{q})\nonumber\\
  &=& 0.
\end{eqnarray}
In passing to the second equality,
we have used $\bar{X}^{(0)}_{,\,a}(\hat{q}) = 0$ proved in Eq. (\ref{eq:app027}).
We have then used Eqs. (\ref{eq:app020}) and (\ref{eq:app022})
in the first and third terms,
respectively, to make these vanish,
whereas the second term vanishes because $\bar{X}^{(0)}_{,\,1 a}(\hat{q}) = 0$
which can be derived from $\bar{X}^{(0)}_{,\, a}(\hat{q}) = 0$.

Then,
we rewrite Eq. (\ref{eq:app026}) as 
\begin{eqnarray}
  \label{eq:app029}
  B^{\alpha\beta}\,X^{(\sigma)}_{,\,\beta}(\hat{\xi})
  = \bar{X}^{(\sigma)}_{,\,1}(\hat{q})\,
  g^\alpha_{,\,1}(\hat{q})
  \,
  \bar{B}^{11}(\hat{q}),\,\,\,\sigma=0,\,1.
\end{eqnarray}
Here,
we have used relations
\begin{eqnarray}
  \label{eq:app030}
  X^{(\sigma)}_{,\,\alpha}(\hat{\xi})
  = \bar{X}^{(\sigma)}_{,\,\mu}(\hat{q})\,f^\mu_{,\,\alpha}(\hat{\xi})
  = \bar{X}^{(\sigma)}_{,\,1}(\hat{q})\,f^1_{,\,\alpha}(\hat{\xi}),\\
  \label{eq:app031}
  B^{\alpha\beta} \, f^1_{,\,\beta}(\hat{\xi})
  = g^\alpha_{,\,\mu}(\hat{q}) \, \bar{B}^{\mu 1}(\hat{q})
  = g^\alpha_{,\,1}(\hat{q}) \, \bar{B}^{11}(\hat{q}),
\end{eqnarray}
which have been derived from
the chain rule and Eqs. (\ref{eq:app005}), (\ref{eq:app006}), (\ref{eq:app010}), and (\ref{eq:app024})-(\ref{eq:app026}).
We note that
Eq. (\ref{eq:app029})
is an alternative form of Eqs. (\ref{eq:app020})-(\ref{eq:app022})
because Eq. (\ref{eq:app029}) can be derived only if all of Eqs. (\ref{eq:app020})-(\ref{eq:app022}) are used.

Substituting the explicit forms
\begin{eqnarray}
  \label{eq:app032}
  X^{(0)}_{,\,\alpha}(\hat{\xi}) &=& V_{,\,\alpha}(\hat{\xi}),\\
  \label{eq:app033}
  X^{(1)}_{,\,\alpha}(\hat{\xi}) &=& 2\,V_{,\,\alpha\beta}(\hat{\xi}) \,
  B^{\beta\gamma} \, V_{,\,\gamma}(\hat{\xi}),
\end{eqnarray}
into Eq. (\ref{eq:app029}),
we have
\begin{eqnarray}
  \label{eq:app034}
  B^{\alpha\beta}\,V_{,\,\beta}(\hat{\xi}) = \gamma(\hat{q}) \, g^\alpha_{,\,1}(\hat{q}),\\
  \label{eq:app035}
  \Big[ B^{\alpha\gamma}\,V_{,\,\gamma\beta}(\hat{\xi}) \,
  - \lambda(\hat{q}) \delta^\alpha_{\,\,\,\beta} \Big] \, g^\beta_{,\,1}(\hat{q}) = 0,
\end{eqnarray}
with
\begin{eqnarray}
  \label{eq:app036}
  \gamma(\hat{q}) &\equiv& \bar{X}^{(0)}_{,\,1}(\hat{q}) \, \bar{B}^{11}(\hat{q}),\\
  \label{eq:app038}
  \lambda(\hat{q}) &\equiv&
  \frac{\bar{X}^{(1)}_{,\,1}(\hat{q})}{2\,\bar{X}^{(0)}_{,\,1}(\hat{q})}.
\end{eqnarray}
Owing to a scale indefiniteness in the definition of $q^1$,
there exists an ambiguity for the overall amplitude of $g^\alpha_{,\,1}(\hat{q})$.
Without loss of generality,
we can adopt the normalization
\begin{eqnarray}
  \label{eq:app039}
  g^\alpha_{,\,1}(\hat{q})\,
  B^{-1}_{\alpha\beta}\,
  g^\beta_{,\,1}(\hat{q}) = 1.
\end{eqnarray}
We note that
Eqs. (\ref{eq:app023}), (\ref{eq:app034}), (\ref{eq:app035}), and (\ref{eq:app039})
are identical to the decoupling conditions.
By setting
\begin{eqnarray}
  \xi^{\alpha=i} &=& x_i,\\
  g^{\alpha=i}_{,\,1}(\hat{q}) &=& \varphi_i/\sqrt{m_i},\\
  q^1 &=& q,
\end{eqnarray}
we find that
Eqs. (\ref{eq:app023}), (\ref{eq:app034}), (\ref{eq:app035}), and (\ref{eq:app039})
correspond to
Eqs. (\ref{eq:002}), (\ref{eq:004}), (\ref{eq:003}), and (\ref{eq:005}),
respectively.

\section{
  Barrier and gate potential with less symmetry
}
\label{sec:008}
A Hamiltonian we consider is given by
\begin{eqnarray}
  H = \frac{1}{2}\,(p_1^2 + p_2^2) + V^\prime_{\mathrm{BG}}(x_1,\,x_2).
\end{eqnarray}
Here, $V^\prime_{\mathrm{BG}}(x_1,\,x_2)$
denotes the barrier and gate potential with less symmetry defined by
\begin{eqnarray}
  V^\prime_{\mathrm{BG}}(x_1,\,x_2)
  = V_{\mathrm{BG}}(x_1,\,x_2)
  - \frac{1}{3}\,\epsilon\,x^3_1,
\end{eqnarray}
where $V_{\mathrm{BG}}(x_1,\,x_2)$ has been given in Eq. (\ref{eq:BG}).
We note that
the term associated with $\epsilon$ breaks the symmetry with respect to $x_1 \rightarrow - x_1$. 
This potential energy surface has two local minima and one saddle point
located at 
$(\epsilon/2b \pm \sqrt{\epsilon^2/4b^2 + a/b},\,
c\epsilon/2\,b \pm c\sqrt{\epsilon^2/4b^2 + a/b})
\equiv (x_{1\pm}^{\mathrm{eq}},\,x_{2\pm}^{\mathrm{eq}})$
and $(0,\,0)$, respectively,
and the activation energies in the transitions
from $(x_{1+}^{\mathrm{eq}},\,x_{2+}^{\mathrm{eq}})$
to $(x_{1-}^{\mathrm{eq}},\,x_{2-}^{\mathrm{eq}})$
and
from $(x_{1-}^{\mathrm{eq}},\,x_{2-}^{\mathrm{eq}})$
to $(x_{1+}^{\mathrm{eq}},\,x_{2+}^{\mathrm{eq}})$
are given by
$a^2/4b + a\epsilon^2/4b^2 + \epsilon^4/24b^3 \pm (a\epsilon/3b+\epsilon^3/12b^2)\sqrt{\epsilon^2/4b^2+a/b}$,
respectively.

\begin{figure}[Hptb]
  \begin{center}
    \begin{minipage}{1.0\linewidth}
      \begin{center}
        \includegraphics[width=0.5\linewidth]{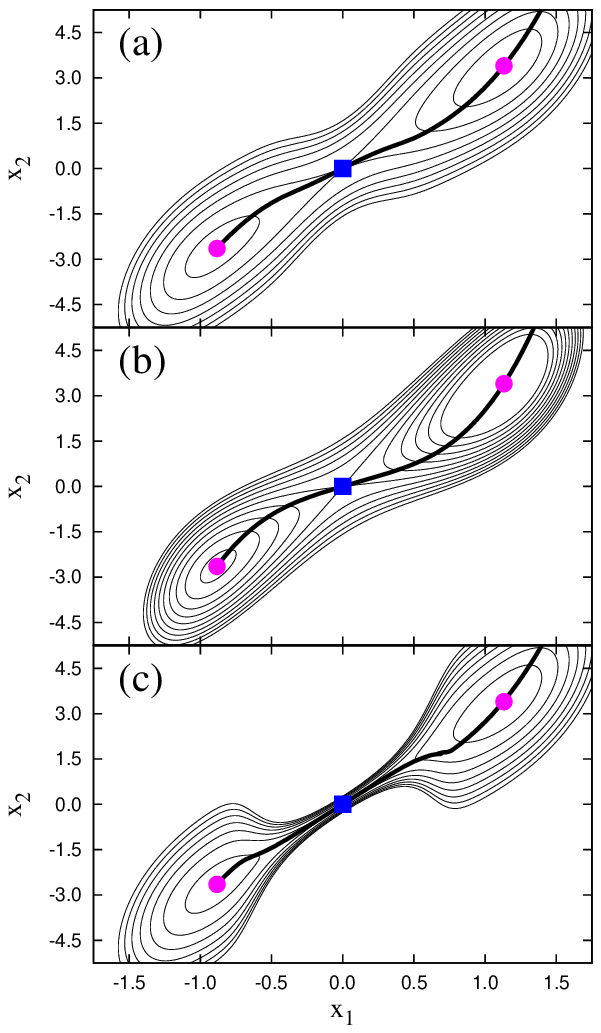}
      \end{center}
    \end{minipage}
  \end{center}
  \caption{
    \label{fig:bgpotential2}
    The collective paths obtained by
    the method developed in this work
    on the $V_{\mathrm{BG}}^{\prime(a)}$, $V_{\mathrm{BG}}^{\prime(b)}$, and $V_{\mathrm{BG}}^{\prime(c)}$ potential energy surfaces.
    The solid lines in (a) ((b) or (c)) show
    the collective path derived from 
    $V_{\mathrm{BG}}^{\prime(a)}$ ($V_{\mathrm{BG}}^{\prime(b)}$ or $V_{\mathrm{BG}}^{\prime(c)}$), respectively.
    In (a) ((b) or (c)),
    the thin lines denote the contours of $V_{\mathrm{BG}}^{\prime(a)}$ ($V_{\mathrm{BG}}^{\prime(b)}$ or $V_{\mathrm{BG}}^{\prime(c)}$),
    the circles the local minima,
    and the squares the saddle point.
  }
\end{figure}

We introduce three potential energy surfaces:
a potential energy surface containing neither a high energy barrier nor a narrow gate $V_{\mathrm{BG}}^{\prime(a)}$,
a surface with a high energy barrier $V_{\mathrm{BG}}^{\prime(b)}$,
and a surface with a narrow gate $V_{\mathrm{BG}}^{\prime(c)}$.
The parameters
$(a,\,b,\,c,\,\alpha,\,\beta,\,\gamma,\,\delta,\,\epsilon)$
which create $V_{\mathrm{BG}}^{\prime(a)}$, $V_{\mathrm{BG}}^{\prime(b)}$, and $V_{\mathrm{BG}}^{\prime(c)}$
read
$(80,\,80,\,3,\,0.0045,\,10,\,10,\,1,\,20)$,
$(240,\,240,\,3,\,0,\,0,\,10,\,20,\,60)$,
and $(80,\, 80,\, \break 3,\, 0.31,\, 690,\, 10,\, 9.7,\,20)$, respectively.
All the potential energies have the same locations of the local minima as
$((1\pm\sqrt{65})/8,\,3(1\pm\sqrt{65})/8)$.

In Fig. \ref{fig:bgpotential2},
we show the collective paths
for $V_{\mathrm{BG}}^{\prime(a)}$, $V_{\mathrm{BG}}^{\prime(b)}$, and $V_{\mathrm{BG}}^{\prime(c)}$.
It reveals that 
the collective paths
provide pathways connecting the two local minima.
We stress that
these results are qualitatively the same as those of
the symmetric potentials $V^{(a)}_{\mathrm{BG}}$,
$V^{(b)}_{\mathrm{BG}}$, and $V^{(c)}_{\mathrm{BG}}$ examined in Sec. \ref{sec:004-2}.

\end{document}